\documentstyle[aps,prb,multicol,epsf]{revtex}
\begin{document}
\draft

\title{Anomalous magnetoconductance due to weak
 localization in 2D systems 
with anisotropic scattering: 
computer simulation}
\author{A. V. Germanenko
\thanks{email: Alexander.Germanenko@usu.ru}, 
V. A. Larionova, G. M. Minkov and S. A. Negashev} 
\address{Institute of Physics and Applied Mathematics, Ural State University \\
620083 Ekaterinburg, Russia}
\date{\today}
\maketitle
\widetext
\begin{abstract}
The results of computer simulation of particle motion over the plane with randomly distributed scatters are presented. They are used to analyse the influence of scattering anisotropy on the negative magnetoresistance due to weak localisation. It is shown that the  magnetic field dependence of magnetoresistance in this case can be described by the well known expression, obtained in the diffusion limit for isotropic scatternig, but with the prefactor less than unity and breaking-phase length which differs from the true value.
\end{abstract}


\bigskip
\begin{multicols}{2}
\narrowtext

\section{Introduction}
Quantum correction to the conductivity arises from interference of electron waves scattered along closed trajectories in opposite directions. An external magnetic field applied perpendicular to the 2D layer  destroys the interference and suppresses the quantum correction. This results in anomalous negative magnetoresistance, which is experimentally observed in many 2D systems. This phenomenon is usually described in the framework of quasiclassical approximation which is justified under the condition $k_F l\gg 1$, where $k_F$ is the Fermi wave vector, $l$ is the mean free path. In this case the conductivity correction is expressed  through the classical probability  density $W$ for an electron to return to the area of the order $\lambda_F l$ around  the start point  \cite{gork,chak,dyak}
\begin{eqnarray}
\label{eq1}
\delta\sigma=-\sigma_0 \frac{\lambda_F l}{\pi} W, \\ \sigma_0=\frac{e^2}{2\pi\hbar} k_F l.
\end{eqnarray}
This expression allows to calculate the conductivity correction at an arbitrary magnetic field \cite{schm}. Analitical expressions are derived only for some specific cases: (i) in the diffusion limit, i.e. at 
$B\ll B_{tr}$ and $l\ll l_\varphi$\cite{gork,chak}, where $l_\varphi$ is the phase-breaking length due to inelastic processes, and $B_{tr}=\Phi_0/2\pi l^2$, $\Phi_0=\pi\hbar c /e$;
(ii) in the high-field limit, i.e. at $B\gg B_{tr}$\cite{dyak}.

To our knowledge, this problem has been solved only for the case of random distribution of scattering centers and isotropic scattering. As a rule, these  conditions are not fulfilled in real semiconductor structures. First of all, in semiconductors the scattering by ionized impurities dominates at low temperatures. This scattering is strongly anisotropic, in particular in heterostructure with a remote doping layer. Besides, the impurity distribution is correlated to some extend due to Coulomb repulsion of the impurity ions at growing temperatures. In the present work the role of 
scattering anisotropy in the quantum  correction to the conductivity is investigated through a computer simulation.

\section{Simulation details}
\begin{figure}
\epsfxsize=\linewidth
\epsfbox {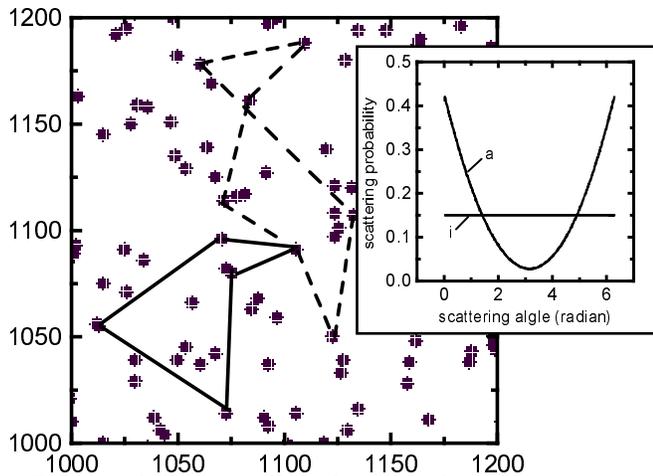}
\caption{\label{f1}
A fragment of the lattice with scattering centers. Lines show closed trajectories. In the inset  the angle dependences of scattering probability for the isotropic (curves {\em i}) and anisotropic (curves {\em a}) scattering mechanisms are presented.}
\end{figure}

To find the values of the mean free path $l$ and probability  density $W$ in (\ref{eq1}), we have simulated the motion of a particle in the 2D plane with scattering centers in it. The plane is a $3000\times3000$ lattice. The scatters are randomly distributed in the lattice sites. The whole scatters number is $2\times 10^4$. Every scatter covers seven lattice parameters in diameter. The start point of  particle motion is chosen in a scatter near the centre of the lattice. We suppose the particle to move linearly between two sequential collisions with scatters (Fig.\ \ref{f1}). The collisions change the motion direction according to given angle dependence of scattering probability. A trajectory is considered to be closed if after a number of collisions $n<n_{max}$ it passes near the start point at the distance less than $d/2$. Since $d$ has to be small enough, we choose $d$ of about the scatter diameter. We assume that the particle never returns to the area of the start point  at $n>n_{max}=1000$ or when it escapes the lattice. An estimation shows that both assumptions  introduce the error in $\delta\sigma$ less than one percent at $\gamma=l/l_\varphi>0.01$. 

The simulation has been carried out for two different scattering mechanisms: isotropic and anisotropic. For the isotropic scattering corresponding to a short-range scattering potential, the scattering probability does not depend on the scattering angle (the curve {\em i} in the inset in Fig.\ \ref{f1}). For the anisotropic scattering we have used the angle dependence of scattering probability presented by the curve {\em a}. It is close to that in heterostructures with a doped barrier, where  impurities are spaced from the 2D gas. The curve {\em a} corresponds, for example, to the 2D structure with impurity density of about $10^{12}$ $cm^{-2}$ and a spacer of $50$ \AA  \ thick.

To calculate the conductivity correction in a magnetic field, we have modified the expression (\ref{eq1}) by including the magnetic field in it according to a standart procedure. The final expression for the conductivity correction in the magnetic field can be written as:

\begin{equation}
\label{eq2}
\frac{\delta\sigma(b)}{G_0}=-\frac{2\pi l}{d\cdot N}\sum_{i}
\cos\left(\frac{bS_i}{l^2}\right)
\exp
\left(-
\frac{l_i}{l_\varphi}
\right),
\end{equation}
where summation runs over all closed trajectories among a total number of trajectories $N$ ($N=10^6$ in our calculations),  $b=B/B_{tr}$,  $G_0=e^2/2\pi^2\hbar$, $S_i$ and $l_i$ stand for the area and length of the $i$-th trajectory, respectively,  the exponent accounts for the phase breaking. Note that in (2) $l$ denotes the mean free path connected with the transport relaxation time. 

\section{Results and discussion}

\begin{figure}
\epsfxsize=\linewidth
\epsfbox {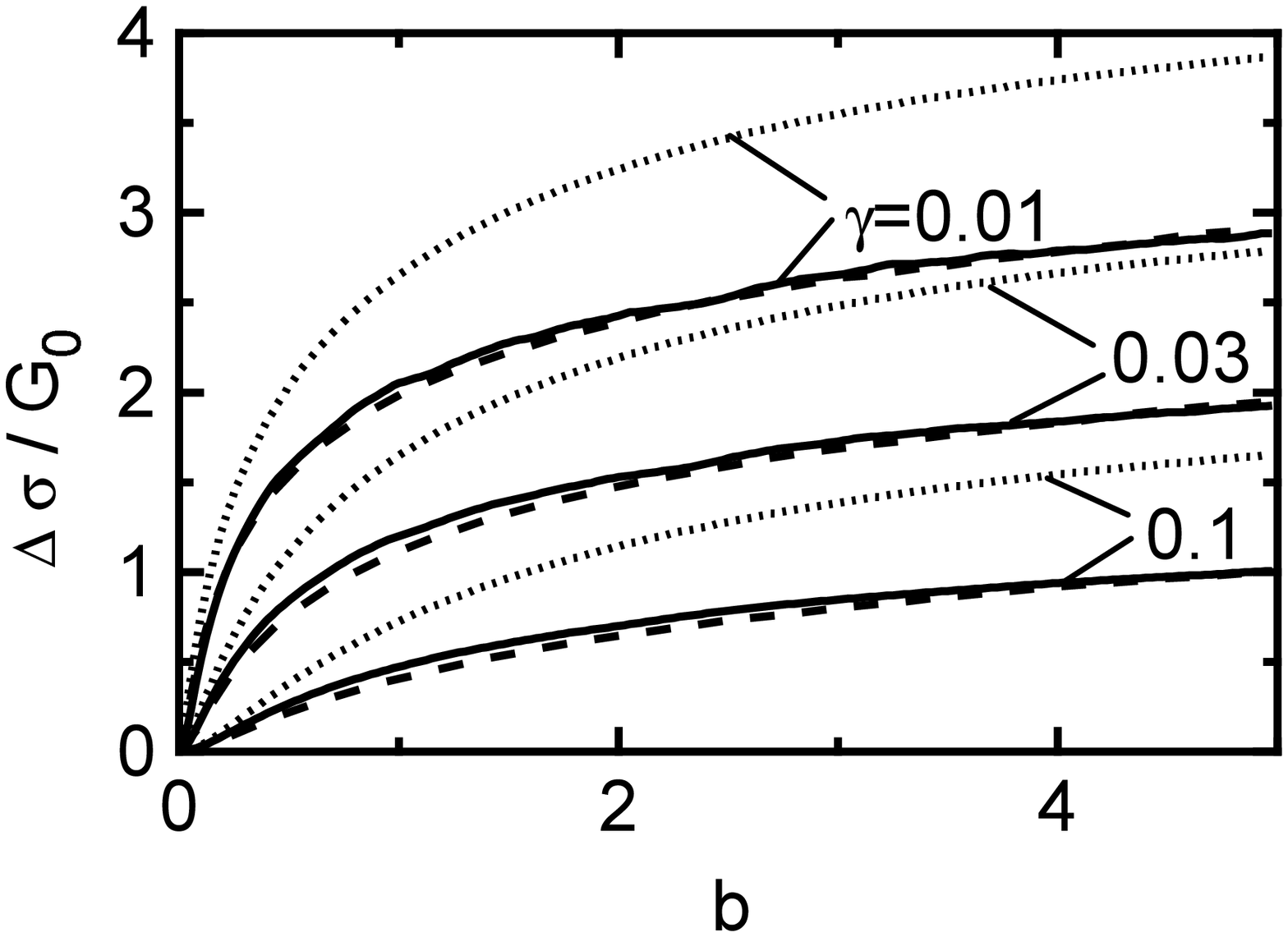}
\caption{
The dependences $\Delta\sigma(b)=\delta\sigma(b)-\delta\sigma(0)$ for the isotropic (dashed curves) and anisotropic (solid curves) scattering mechanisms, which are calculated with the angle dependences of scattering probabilities presented in the inset in Fig.\ \protect\ref{f1}. Dotted curves are the results of diffusion approximation (\protect\ref{eq3}).}
\label{f2}
\end{figure}

The results of simulation, obtained for different $\gamma$ values, are presented in Fig.\ \ref{f2}. Let us consider, at first, the results for the isotropic scattering (dashed curves). They are in a good agreement with the results of numerical calculation carried out beyond the diffusion limit \cite{schm}. This lends support to the validity of the method used. Here the results of calculation in the diffusion limit ($b\ll 1$, $\gamma\ll 1$), obtained through the well-known expression 
\begin{equation}
\label{eq3}
\frac{\Delta\sigma(b)}{G_0}=
\psi\left(\frac{1}{2}+\frac{\gamma}{b}\right)-
\psi\left(\frac{1}{2}+\frac{1}{b}\right)-
\ln(\gamma),
\end{equation}
are shown too. As is clearly seen, the conditions, under which the expression (\ref{eq3}) works well, are very rigorous: even for $\gamma=0.01$ and $b<1$ the formula (\ref{eq3}) gives evidently larger value of $\Delta\sigma(b)$ than that in our simulation. 
\begin{figure}
\epsfxsize=\linewidth
\epsfbox {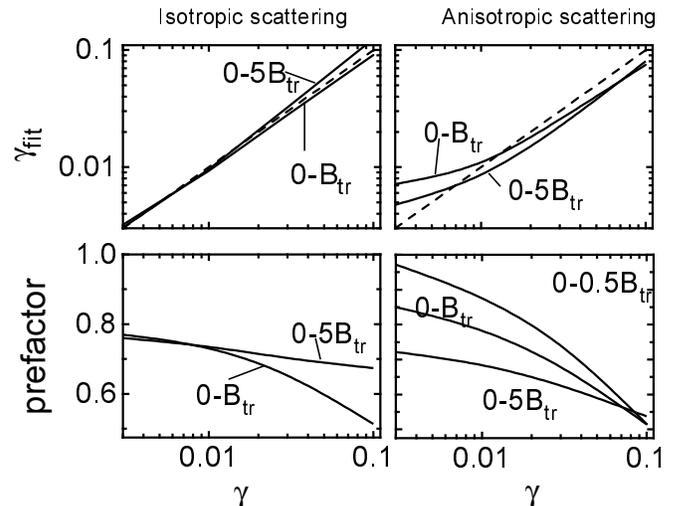}
\caption{
The results of  fitting of simulated curves (solid and dashed curves in Fig.\ \protect\ref{f1}) by expression (\protect\ref{eq3}). Dashed lines in upper figures correspond to $\gamma_{fit}=\gamma$.}
\label{f3}
\end{figure}

The solid curves in Fig.\ \ref{f2} are the results of our simulation for the anisotropic scattering mechanism. The closeness of these results to those, obtained for isotropic scattering, stands out. At first glane it seems to be surpising. The introduction of anisotropy in scattering process decreases the probability of returning to the start point area,  because the scattering for small angles dominates in this case and in the average the particle moves to greater  distance from the start point after each collision. As is seen from (\ref{eq1}), the decreasing of $W$ has to lead to decreasing in $\delta\sigma$. However, this decreasing is compensated by the increasing of transport length $l$, which is averaged with the weight $(1-\cos(\theta))$ when integration runs over the scattering angle $\theta$. Thus, the introduction of anisotropy of scattering does not dramatically change the negative magnetoresistance (at least up to $b=5$). 

One can now attack the results of simulation as experimental ones. Let us describe our curves by expression (\ref{eq3}) in the usual fashion. We have introduced a numerical multiplier (so-called prefactor) in (\ref{eq3}) and used it and $\gamma$ as fitting parameters. The results of such data processing are presented in Fig.\ \ref{f3}. The curves labels show the magnetic field ranges, in which the fitting procedure has been carried out. In spite of strong difference between the simulation results and those given by (\ref{eq3}) (see dotted and other curves in Fig.\ \ref{f2}), the fitting value of $\gamma_{fit}$ is very close to $\gamma$, used in our simulation for both isotropic and anisotropic scattering mechanisms. The difference between $\gamma$ and $\gamma_{fit}$ is less than $10$ \% for the isotropic scattering and $40-50$ \% for the anisotropic one. The value of prefactor  is less than unity and decreases with increasing $\gamma$. 

Thus, (i) the use of expression (\ref{eq3}) to determine the phase-breaking length (or time) from the magnetic field dependence of anomalous magnetoconductance in semiconductor structures, where scattering is anisotropic, can give the error of about $40-50$ \%, (ii) the fact that the prefactor is less than $1$ can be related not only with e-e interaction influence as it is frequently supposed, but with poor fulfilment of the condition $l\ll l_\varphi$ and  strong anisotropy of scattering as well.

\subsection*{Acknowledgements} 
This work was supported in part by the RFBR through Grants  
97-02-16168,  98-02-17286, the Russian Program 
{\it Physics of Solid State Nanostructures} through Grant  97-1091,
and the Program {\it University of Russia} through Grant 420.

\end{multicols}
\end{document}